%% file: main.tex
\documentclass[sigconf]{acmart}
\usepackage{multirow}
\AtBeginDocument{%
  \providecommand\BibTeX{{%
    \normalfont B\kern-0.5em{\scshape i\kern-0.25em b}\kern-0.8em\TeX}}}

\setcopyright{none}
\renewcommand\footnotetextcopyrightpermission[1]{}
\begin{document}

\title[Behavioral Forensics in Social Networks]{Behavioral Forensics in Social Networks: Identifying Misinformation, Disinformation and Refutation Spreaders Using Machine Learning}

\author{Euna Mehnaz Khan$^{*, 1}$, Ayush Ram$^2$, Bhavtosh Rath$^3$, Emily Vraga$^4$, and Jaideep Srivastava$^5$}
% \authornote{Euna Mehnaz Khan is the corresponding author of this paper.}
% \orcid{0000-0002-0030-7841}
% \affiliation{%
%   \institution{University of Minnesota}
% \city{Minneapolis}
%   \state{MN}
% \email{khan0586@umn.edu}

% \author{Ayush Ram}
% \affiliation{%
%   \institution{University of Minnesota}
% %   \streetaddress{P.O. Box 1212}
% \city{Minneapolis}
%   \state{MN}
% %   \postcode{}
% }
% \email{ram00009@umn.edu}

% \author{Bhavtosh Rath}
% \affiliation{%
%   \institution{University of Minnesota}
% %   \streetaddress{}
%   \city{Minneapolis}
%   \state{MN}
% \email{rathx082@umn.edu}

% \author{Emily Vraga}
% \affiliation{%
%   \institution{University of Minnesota}
% %   \streetaddress{P.O. Box 1212}
%   \city{Minneapolis}
%   \state{MN}
% %   \postcode{}
% }
% \email{ekvraga@umn.edu}

% \author{Jaideep Srivastava}
% \affiliation{%
%   \institution{University of Minnesota}
% %   \streetaddress{P.O. Box 1212}
%   \city{Minneapolis}
%   \state{MN}
% %   \postcode{}
% }
% \email{srivasta@umn.edu}
\renewcommand{\shortauthors}{Khan, et al.}
\thanks{This work was first submitted to CIKM 2022. \\
$^*$Euna Mehnaz Khan is the corresponding author of this paper. All the authors are with the University of Minnesota, Minneapolis, MN. Email: {\tt \{$^1$khan0586, $^2$ram00009, $^3$rathx082, $^4$ekvraga, $^5$srivasta\}@umn.edu}}
\begin{abstract}
  \input{sections/abstract}

\end{abstract}





\maketitle
\pagestyle{empty}

\input{sections/intro}

\input{sections/rel_work}

\input{sections/behavioral_forensics}

\input{sections/method}
\input{sections/dataset}

\input{sections/experiment}

\input{sections/discussion.tex}

\input{sections/conclusion}

 \section*{Acknowledgements}
 We would like to acknowledge the Minnesota Supercomputing Institute (MSI) at the University of Minnesota for providing resources that contributed to the research results reported within this paper. URL: \url{http://www.msi.umn.edu}

\bibliographystyle{ACM-Reference-Format}
\bibliography{mybibfile}

\end{document}

%% file: sections/abstract.tex
% background-motivation of the research, need for new research.
% method, contribution
% results
%---------------abstract---------------
With the ever-increasing spread of misinformation on online social networks, it has become very important to identify the spreaders of misinformation (unintentional), disinformation (intentional), and misinformation refutation. It can help in educating the first, stopping the second, and soliciting the help of the third category, respectively, in the overall effort to counter misinformation spread. Existing research to identify spreaders is limited to binary classification (true vs false information spreaders). However, people's intention (whether naive or malicious) behind sharing misinformation can only be understood after observing their behavior after exposure to both the misinformation and its refutation which the existing literature lacks to consider. In this paper, we propose a labeling mechanism to label people as one of the five defined categories based on the behavioral actions they exhibit when exposed to misinformation and its refutation. However, everyone does not show behavioral actions but is part of a network. Therefore, we use their network features, extracted through deep learning-based graph embedding models, to train a machine learning model for the prediction of the classes. We name our approach \textit{behavioral forensics} since it is an evidence-based investigation of suspicious behavior which is spreading misinformation and disinformation in our case. After evaluating our proposed model on a real-world Twitter dataset, we achieved 77.45\% precision and 75.80\% recall in detecting the \textit{malicious} actors, who shared the misinformation even after receiving its refutation. Such behavior shows intention, and hence these actors can rightfully be called agents of disinformation spread.

%% file: sections/intro.tex
% background/motivation, application  
% current knowledge
% Limitations of Existing Approaches
% proposed solution, contribution
% experimental design, evaluation, results, comparison of results to existing works
% organization of the paper (optional)
%---------------------------
\section{Introduction}
%misinformation spread in OSNs
In recent times, the ease of access to Online Social Networks and the extensive reliance on it for news has increased the dissemination of misinformation. The spread of misinformation has severe impacts on our lives which has also been witnessed during the COVID-19 pandemic~\cite{banerjee2021covid}. Hence, it is important to detect misinformation along with its spreaders. It is worth to be noted that misinformation and disinformation are very related yet different terms. Misinformation is incorrect or misleading information~\cite{mw:misinformation} whereas disinformation is spread deliberately with the intention to deceive~\cite{mw:disinformation}.

\input{fig_tab_tex/method_overview}

% witnessed during the COVID-19 pandemic~\cite{banerjee2021covid}

%motivation and application
Fact-checking websites often debunk misinformation and publish its refutation. As a result, both the misinformation and its refutation can co-exist in the network and people can be exposed to them in different orders. At some point in time, they might get exposed to the misinformation, and retweet it. Later, they may get exposed to its refutation and retweet it. Since they have corrected their mistake, it can be inferred that they had spread the misinformation unintentionally. However, social media usually bans or flags accounts that they deem to be objectionable~\cite{wired2021news} without investigating the intention of those people sharing the misinformation. This results in many unfair bans of accounts that were simply deceived by the misinformation. On the other hand, some people may not correct their mistakes, or despite receiving the refutation they may choose to retweet the misinformation. These kinds of activities reveal their bad intention and hence these people can be considered as \textit{malicious}. Again, some people might be smart enough to identify misinformation and choose to share refutations instead which indicates good intentions. Identifying these different groups of people will enable us to mitigate misinformation spread efficiently. For instance, Twitter may flag or ban the \textit{malicious} people besides incentivizing the good people to spread the refutation. The followers of \textit{malicious} people can also be treated with the refutation as a preventive measure~\cite{ozturk2015combating}. Inoculating people against common misinformation and misleading tactics has shown promise, so targeting vulnerable groups more precisely offers great advantages~\cite{lewandowsky2021countering}. In this paper, we build a model to identify these different groups of spreaders using their behavioral and network properties. The novelty of our approach is that it considers the possible series of actions (retweet or not retweet) a person might take when exposed to both misinformation and its refutation.
% refutation as a preventive measure~\cite{ozturk2015combating}

% current knowledge and limitations of Existing Approaches
Most research on misinformation and refutation spreader detection utilizes users' profile features (i.e., gender, location, follower-followee count, linguistic features, sentiment extracted from content posted by them)~\cite{wang2019machine, bodaghi2020characteristics}. Some research~\cite{rath2020detecting, tommasel2022following} has considered network features extracted through graph embedding algorithms to identify the spreaders. However, these works considered only two classes of spreaders (misinformation and refutation spreaders). In these works, people are labeled as misinformation spreaders if they retweet that information just once, despite the fact that given the refutation, they might have acted differently. On the other hand, if a person shares the refutation, they are considered true information spreaders whereas they might have shared the misinformation if received earlier, in which case the same model identifies them as misinformation spreaders. Thus, only part of the whole scenario has been considered in existing literature. 

% \input{fig_tab_tex/method_overview}

%proposed solution and results
In this paper, we build a model to classify different kinds of spreaders. First, we label people into one of our five defined classes. Our labeling mechanism is novel because we label people only after they have been exposed to both misinformation and its refutation aiming to understand their intention and consider multiple levels of good and bad behavior. However, it is not possible to classify everyone solely based on behavior because everyone does not have behavioral data. But everyone is part of the social network. Hence, features from their networks are used to predict the categories of people without behavioral histories. For that purpose, from the follower-followee network of these labeled people, we extract their network features using graph embedding models, like Large-scale Information Network Embedding (LINE)~\cite{tang2015line} and PyTorch-BigGraph~\cite{lerer2019pytorch}. Then, we use these learned network features along with the profile features of the labeled people to train a machine learning classification model and predict the labels. In this way, for users without past behavioral histories, we can predict the labels from their network and profile features. We have tested our model on a Twitter dataset, and have achieved 77.45\% precision and 75.80\% recall in detecting the \textit{malicious} class (the extreme bad) where the accuracy of the model is 73.64\% with a weighted F1 score of 72.22\%, thus significantly outperforming the baseline models. The contributions of this paper are as follows:
\begin{itemize}
    \item We observe people's sequence of behaviors when they are exposed to false information and its refutation, and label them only after they have been exposed to both.
    \item We propose a nuanced labeling mechanism that uses behavioral data to label people into five categories. This is a granular level classification of misinformation, disinformation, and refutation spreaders which aims to identify the intent of people based on their history of actions.
    \item We leverage the network embedding features of users to predict their classes so that people without any behavioral property can also be categorized.
\end{itemize}
An overview of the proposed approach, which we name \textit {behavioral forensics}, is demonstrated in Fig.~\ref{fig:overview}.
% We name our approach \textit{behavioral forensics} since the word, `forensic' means evidence-based investigation of suspicious behavior, typically criminal behavior and we are doing the same using behavioral actions in the context of misinformation spread.

%% file: fig_tab_tex/method_overview.tex
\begin{figure*}[t]
  \centering
  \includegraphics[width=.95\linewidth]{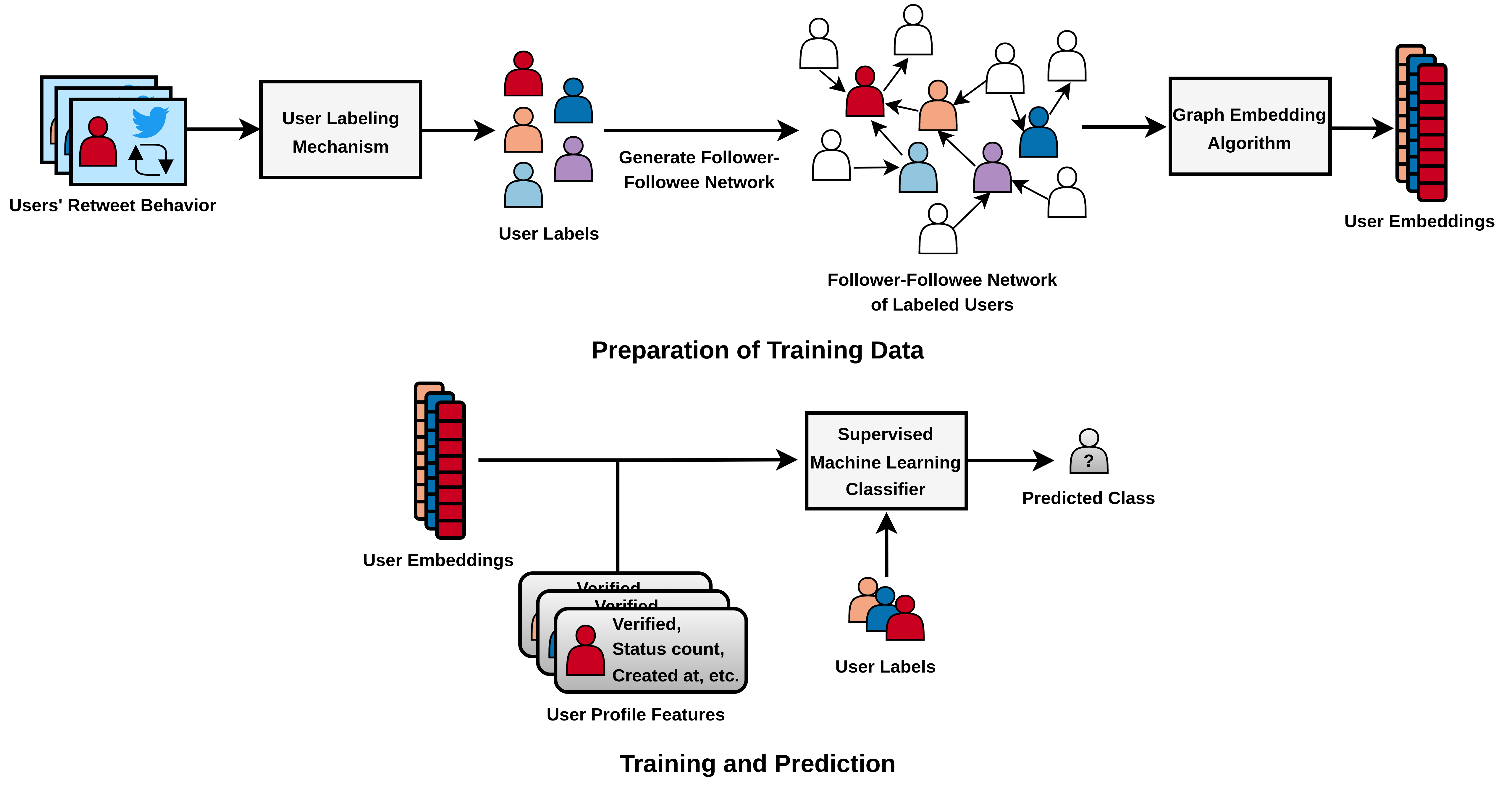}
  \caption{Overview of the proposed \textit{behavioral forensics} model which generates labels from people's retweet behavior and utilizes both network and profile features to train machine learning classifiers.}
%   \Description{Describe the diagram.}
  % \vspace{-4mm}
  \label{fig:overview}
\end{figure*}

%% file: sections/rel_work.tex
\section{Related Work}
\label{sec-related}
%works on misinformation detection 
Existing research in the area of misinformation detection mostly includes analyzing the content (i.e., linguistic, stylistic features, etc.)~\cite{volkova2018misleading, horne2017just}, using deep learning-based models~\cite{liu2018early, lu2020gcan}, propagation structure of misinformation in the network~\cite{ma2017detect}, users' profile features (account age, number of tweets and followers)~\cite{bodaghi2020characteristics}, their history of sharing false and true information~\cite{shu2018understanding}, etc. Research on identifying spreaders of misinformation and refutations is mostly centered around utilizing users' profile features~\cite{wang2019machine}, linguistic and personality features from timeline tweets~\cite{shrestha2020detecting, giachanou2022impact}, and sentiment and topics from past tweets~\cite{li2020social}. Other works include identifying the misinformation source~\cite{zhu2014information}, finding minimal-size clusters of fake news spreaders~\cite{alassad2019finding} or refutation spreaders~\cite{nguyen2012containment} who can influence the most people. Recently people's interactions and network structures are also being considered while identifying the misinformation spreaders~\cite{rath2020detecting, tommasel2022following, sharma2021identifying, shrestha2022characterizing}. However, none of the past works has considered the fact that a person can exhibit a series of behaviors when exposed to misinformation and its refutation. People's perceptions of truth change as they get more exposed to the facts. They may retract their previous actions (retweeting the misinformation) by doing something opposite (retweeting the refutation) to account for their mistake, which implies good behavior. On the other hand, labeling them as \textit{malicious} or bad people based on the fact that they chose not to share a refutation and instead shared the misinformation gives more evidence than just relying on the fact that they shared the misinformation. In this paper, we have identified the multiple states that one can go through when exposed to both misinformation and its refutation, and classify them using their network properties.

%% file: sections/behavioral_forensics.tex
\section{Behavioral Forensics in Social Networks}
\label{sec-foren}
Every organization has some policy for identifying suspects based on their behavior. For example, many finance organizations will block a user after three failed password attempts, on the suspicion that the attempts are from a potential hacker. A similar approach is taken by organizations in identifying who might be originators and/or spreaders of misinformation, based on their behavior. Such approaches tend to be organization specific, and can be expressed as a set of labeling rules. {\it Our proposed research is not tied to any specific policy for identifying suspected perpetrators, but rather can take any policy and work with it. To illustrate our work, we pick a specific policy, which we believe is reasonable, but does not represent any specific organization. It is merely an illustrative example.} The overall approach is shown in Fig.~\ref{fig:overview}, which requires a labeling policy, but not any specific one.

While labeling people, we consider only those who are exposed to at least one pair of misinformation and its refutation. Next, we observe their sequence of actions upon exposures to those messages and label them into one of the five categories using our proposed definitions described below. The second part (generating the embedding features using the follower-followee network of the labeled people and training machine learning models) is independent of the labeling mechanism (policies used to define spreaders) used. The categories can be predicted using embedding features which is our main contribution. However, the key novelty introduced regarding the labeling here is that people should be labeled by observing their actions only when they have been exposed to both misinformation and its refutation. The underlying policy to define the categories can differ from ours but has to be rational.

{\bf Example policy:} A possible series of behavioral actions is depicted using a state diagram in Fig.~\ref{fig:state_diagram}. Here, misinformation and its refutation are denoted by $r_m$ and $r_f$, respectively. Sharing or retweeting $r_m$ is represented by $\texttt{share}(r_m)$ whereas exposure to $r_m$ is represented by $\texttt{exp}(r_m)$. Same applies for $r_f$. Self-loops are not shown in the diagram. For instance, if a person is at state $\mathbf{G}$ and they share an $r_m$, they go to state $\mathbf{I}$. Now, they stay at state $\mathbf{I}$ if they repeatedly keep sharing the $r_m$. The self-loops are removed from the diagram for clarity of the figure. The definitions used for our proposed five classes are described below:
\input{fig_tab_tex/state_diagram}

\begin{enumerate}
    \item \textbf{\textit{malicious}}: These are the people who spread misinformation knowingly. After being exposed to both $r_m$ and $r_f$, they decide to spread $r_m$, not $r_f$. Since refutations are published by fact-checking websites, they are clear to identify as true and usually are not confused with $r_m$. Therefore, when a person shares the $r_m$ even after getting the $r_f$, they can be considered to have malicious intent, hence categorized as \textit{malicious}.
    
    In Fig.~\ref{fig:state_diagram}, a person in state $G$ or $O$ is exposed to both $r_m$ and $r_f$ ($\mathbf{A}$ $\rightarrow$ $\mathbf{B}$ $\rightarrow$ $\mathbf{G}$, $\mathbf{A}$ $\rightarrow$ $\mathbf{J}$ $\rightarrow$ $\mathbf{O}$). Now, if they choose to share the $r_m$, they go to state $I$ ($\mathbf{G}$ $\rightarrow$ $\mathbf{I}$) and $P$ ($\mathbf{O}$ $\rightarrow$ $\mathbf{P}$), respectively. If they do not take further action, or repeatedly keep sharing the $r_m$, we label them as \textit{malicious}. In another scenario, a person shares $r_m$ after getting it ($\mathbf{A}$ $\rightarrow$ $\mathbf{B}$ $\rightarrow$ $\mathbf{C}$) and then they get the $r_f$ ($\mathbf{C}$ $\rightarrow$ $\mathbf{D}$). Now, if they still choose to share the $r_m$ again ($\mathbf{D}$ $\rightarrow$ $\mathbf{F}$), we label them as \textit{malicious}. The \textit{malicious} class of people is an extremely bad category in terms of misinformation spread and should be banned or flagged.

    \item \textbf{\textit{maybe\_malicious}}: This class refers to the following groups of people:
    \begin{itemize}
        \item people who have shared the misinformation ($\mathbf{A}$ $\rightarrow$ $\mathbf{B}$ $\rightarrow$ $\mathbf{C}$), then have received its refutation ($\mathbf{C}$ $\rightarrow$ $\mathbf{D}$) but did not share the refutation (they stay at state $\mathbf{D}$). These people did not correct their mistake which may indicate malicious intent or it is possible that they got the $r_f$ so late that the topic has become outdated. They are not necessarily the bad group of people. We can keep an eye on them and monitor their activities. 
        \item people who shared the $r_m$ after sharing its $r_f$ (perhaps they cannot distinguish between true and false information although $r_f$s are usually clear to identify as true). The sequences $\mathbf{A}$ $\rightarrow$ $\mathbf{B}$ $\rightarrow$ $\mathbf{G}$ $\rightarrow$ $\mathbf{H}$ $\rightarrow$ $\mathbf{S}$, $\mathbf{A}$ $\rightarrow$ $\mathbf{J}$ $\rightarrow$ $\mathbf{K}$ $\rightarrow$ $\mathbf{L}$ $\rightarrow$ $\mathbf{M}$, $\mathbf{A}$ $\rightarrow$ $\mathbf{J}$ $\rightarrow$ $\mathbf{O}$ $\rightarrow$ $\mathbf{Q}$ $\rightarrow$ $\mathbf{U}$ demonstrate this kind of behavior.
    \end{itemize}
    The intent of these two subgroups of people is not clear from their behavior. But since they shared $r_m$ as the latest action, we grouped them as a separate category called \textit{maybe\_malicious}. These people are not as bad as the \textit{malicious} class of people, however, they still contribute to misinformation spread. Less intense measures like providing refutations to their followers can be taken to account for the harm caused by them.
    
    \item \textbf{\textit{naive\_self\_corrector}}: These people got deceived by the $r_m$ and shared it (naive behavior) but later corrected their mistake by sharing the $r_f$ (self-correcting behavior). The sequences $\mathbf{A}$ $\rightarrow$ $\mathbf{B}$ $\rightarrow$ $\mathbf{C}$ $\rightarrow$ $\mathbf{D}$ $\rightarrow$ $\mathbf{E}$, $\mathbf{A}$ $\rightarrow$ $\mathbf{B}$ $\rightarrow$ $\mathbf{G}$ $\rightarrow$ $\mathbf{I}$ $\rightarrow$ $\mathbf{T}$, $\mathbf{A}$ $\rightarrow$ $\mathbf{J}$ $\rightarrow$ $\mathbf{O}$ $\rightarrow$ $\mathbf{P}$ $\rightarrow$ $\mathbf{R}$ fall into this category. These people can be provided $r_f$ early to prevent them from naively believing and spreading $r_m$ and be utilized to spread the true information.
    
    \item \textbf{\textit{informed\_sharer}}: This category includes two types of people:
    \begin{itemize}
        \item People who shared only $r_f$ after exposure to both $r_m$ and $r_f$ ($\mathbf{A}$  $\rightarrow$ $\mathbf{B}$ $\rightarrow$ $\mathbf{G}$ $\rightarrow$ $\mathbf{H}$, $\mathbf{A}$  $\rightarrow$ $\mathbf{J}$  $\rightarrow$ $\mathbf{O}$  $\rightarrow$ $\mathbf{Q}$).
        \item People who shared $r_f$ ($\mathbf{A}$ $\rightarrow$ $\mathbf{J}$ $\rightarrow$ $\mathbf{K}$), then after receiving $r_m$ ($\mathbf{K}$ $\rightarrow$ $\mathbf{L}$), did not share it (stay at state $L$) or they shared the refutation again ($\mathbf{L}$ $\rightarrow$ $\mathbf{N}$). Both the states $\mathbf{L}$ and $\mathbf{N}$ fall under this category. 
    \end{itemize}
    This group of people are smart enough to distinguish between true and false information and are willing to fight misinformation spread by sharing the refutation. Therefore, they should be provided with refutations on the onset of misinformation dissemination to contain its spread.
    
    \item \textbf{\textit{disengaged}}: People who received both $r_m$ and $r_f$ but shared nothing are defined as \textit{disengaged} people. This group of people does not incline to share any true or false information. People in state $\mathbf{G}$ (sequence $\mathbf{A}$ $\rightarrow$ $\mathbf{B}$ $\rightarrow$ $\mathbf{G}$) and state $\mathbf{O}$ (sequence $\mathbf{A}$ $\rightarrow$ $\mathbf{J}$ $\rightarrow$ $\mathbf{O}$) are \textit{disengaged} people.
\end{enumerate}
It should be noted that when a person is in state $\mathbf{G}$ and takes no further action, they are identified as \textit{disengaged}. But if they share something at this point, then they make a transition to state $\mathbf{H}$ or $\mathbf{I}$ depending on what they share. For instance, if they share $r_f$, they go to state $\mathbf{H}$. Now, if they stop here, then they are defined as \textit{informed\_sharer}. However, if they share $r_m$ here, they go to state $\mathbf{S}$ which indicates \textit{maybe\_malicious} class.

%% file: fig_tab_tex/state_diagram.tex
\begin{figure}[t]
  \centering
  \includegraphics[width=\linewidth]{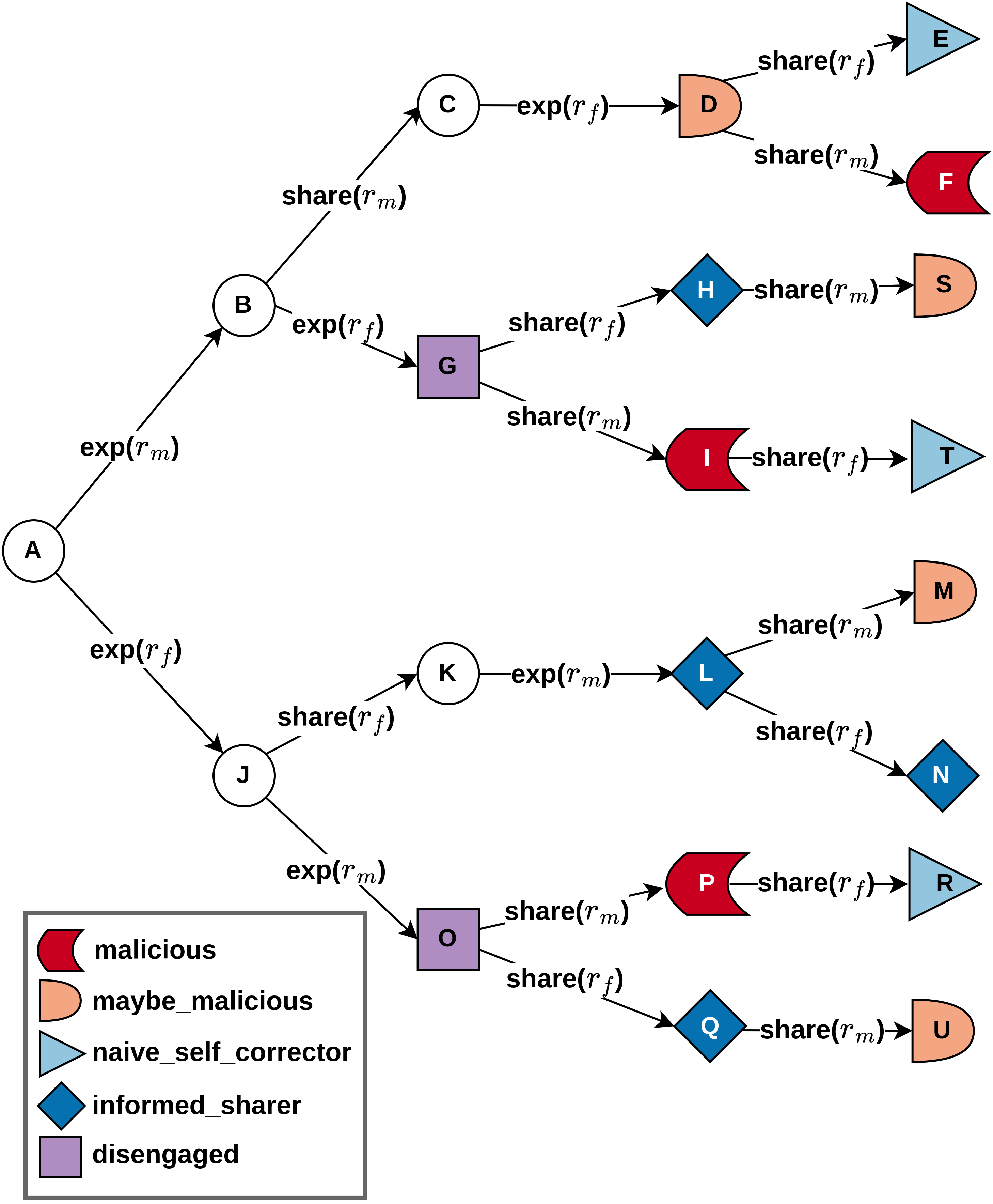}
  \caption{State diagram for the proposed labeling mechanism demonstrating different sequences of behavior upon exposure, \texttt{exp()}, to a piece of misinformation ($r_m$) and its refutation ($r_f$).} 
  %   \Description{Describe the diagram.}
  % \vspace{-4mm}
  \label{fig:state_diagram}
\end{figure}

%% file: sections/method.tex
\section{Proposed Method}
\label{sec-method}
In this paper, our goal is to classify people into one of the five categories defined in Section~\ref{sec-foren}. We follow the methodology described below to accomplish our goal:

\subsection{Labeling Mechanism}
First, we label the people who have been exposed to misinformation and its refutation. These labels will serve as gold standard labels~\cite{enwiki:1110106256} (as opposed to ground truth labels~\cite{enwiki:1092727524}) for training of our machine learning models. 

% Ground truth is information that is directly measurable or observable \cite{enwiki:1092727524}. Here, we are trying to understand and label people's intention which is not directly measurable. In such cases, gold standard label, created by experts by inference and annotation, is used which tries to represent ground truth as closely as possible \cite{enwiki:1110106256}. 

In particular, we have considered multiple pairs of misinformation ($r_m$) and corresponding refutations ($r_f$) for this task (as discussed in Section~\ref{sec-dataset}). Ideally, we wanted to label someone as one class, i.e., \textit{malicious} if they had shown that behavior multiple times across our many pairs of $r_m$ and $r_f$. Although this would have been a more robust labeling of people, we observed during the experiment that a very small number of people showed behavior that falls into a class other than \textit{disengaged}, and the number of people exhibiting that behavior multiple times was even less. To account for this problem, we labeled people according to our state diagram in Fig.~\ref{fig:state_diagram} whenever they got exposed to at least one pair of $r_m$-$r_f$. For users who got exposed to only one pair of $r_m$ and $r_f$, received a single final label. On the other hand, people who got exposed to multiple pairs of $r_m$ and $r_f$, got multiple labels. If a person receives multiple same labels, then we label them as the corresponding class. However, if they receive other label(s) with \textit{disengaged} label(s), then we remove the \textit{disengaged} label(s) and use the rest of the non \textit{disengaged} label(s) to convert them into a single label. We do this because showing \textit{disengaged} behavior is trivial since it means taking no action whereas the other behaviors require some sort of action which is essential to identify. Next, we use a Likert scale-like~\cite{likert1932technique} representation to convert multiple labels for a user into a single label, as described below:

First, we represent the four different non \textit{disengaged} classes (\textit{malicious}, \textit{maybe\_malicious}, \textit{naive\_self\_corrector}, \textit{informed\_sharer}) using the integers 1, 2, 3, and 4, respectively. Suppose, we have $m$ labels for a person represented by the list $l=[l_1, l_2, ..., l_m]$. After converting the labels to the integers, we get a list of integers $[n_1, n_2, ..., n_m]$. Since we are studying behavior, and the distribution of labels for the same user was skewed, we take the median of these numbers, choosing the larger one in case of a tie to avoid false positives (considering the \textit{malicious} as positive). However, if we had used mean (with rounding up to get an integer) instead of median, it would have changed the final label for only 8 users out of 218 multilabel users.

For example, if $l=$[\textit{malicious}, \textit{naive\_self\_corrector}, \textit{informed\_sharer}], then we get 3 as the median of $[1,3,4]$, which refers to the class \textit{naive\_self\_corrector}. 

\subsection{Graph Embeddings} 
\label{sec-emb}
Since very few people exhibit behavioral actions as observed during our experiment, it would not be possible to label everyone using only their behavioral data. In contrast, everyone has network properties since they are part of the social network. Hence, we create a labeled dataset where labels come from the labeling mechanism stated in Section~\ref{sec-foren} and the features come from their network extracted using graph embedding models. This labeled dataset can be used to train a machine learning classifier to predict the category of a person given their network embedding features.

Graph embedding models are used to generate a low dimensional vector representation for each of the nodes in a network, preserving the network's topology and homophily of the nodes. Nodes with similar neighborhood structure should have similar vector representations. As we aim to utilize network properties of the people to distinguish between different classes, we apply the existing graph embedding methods. In particular, after labeling people, we build a network using the followers and followees of the labeled users. Then, we use a graph embedding model to extract the network features of these users. Specifically, we have tried second order version of LINE~\cite{tang2015line} (as the network is directed) and PyTorch-BigGraph (PBG)~\cite{lerer2019pytorch} for this purpose. LINE algorithm captures the local and global network structure by considering the fact that similarity between two nodes is also dependent on the number of neighbors they share besides the existence of direct link between them. This is important in our problem because people from the same class may not be connected to each other but they might be connected to the same group of people, which the LINE algorithm still identifies as a similarity. For instance, people from the \textit{malicious} class may or may not be connected to each other but their target people (whom they want to relay the misinformation to) might be the same.  Again, nodes from the same class may form a cluster or community with lots of interconnections and common neighbors. LINE graph embedding technique is able to capture these aspects. On the other hand, PBG embedding system uses a graph partitioning scheme
that allows it to train embeddings quickly and scale to networks with millions of nodes and trillions
of edges. Its effectiveness and fast generation of embeddings for social network dataset (Twitter) is also shown in~\cite{lerer2019pytorch}. For this reason, we also try this embedding technique.

% and found their similar ability to distinguish between different classes (details in Section~\ref{sec-exp}). 

\input{fig_tab_tex/class_overview}

\subsection{Node Classification}
In this step, we combine the profile features of the users with their learned embeddings to train different machine learning models (more details in Section~\ref{sec:exp-classification}). Finally, we use these models to make predictions. 
Due to the heavy imbalance between the \textit{disengaged} class and other classes, we perform the classification in two steps as shown in Fig.~\ref{fig:class_overview}. First, we classify people into the \textit{disengaged} category and \textit{others} category with undersampling of the \textit{disengaged} class. Next, we classify the \textit{others} category into the four defined classes.

%% file: fig_tab_tex/class_overview.tex
\begin{figure}[t]
  \centering
  \includegraphics[width=\linewidth]{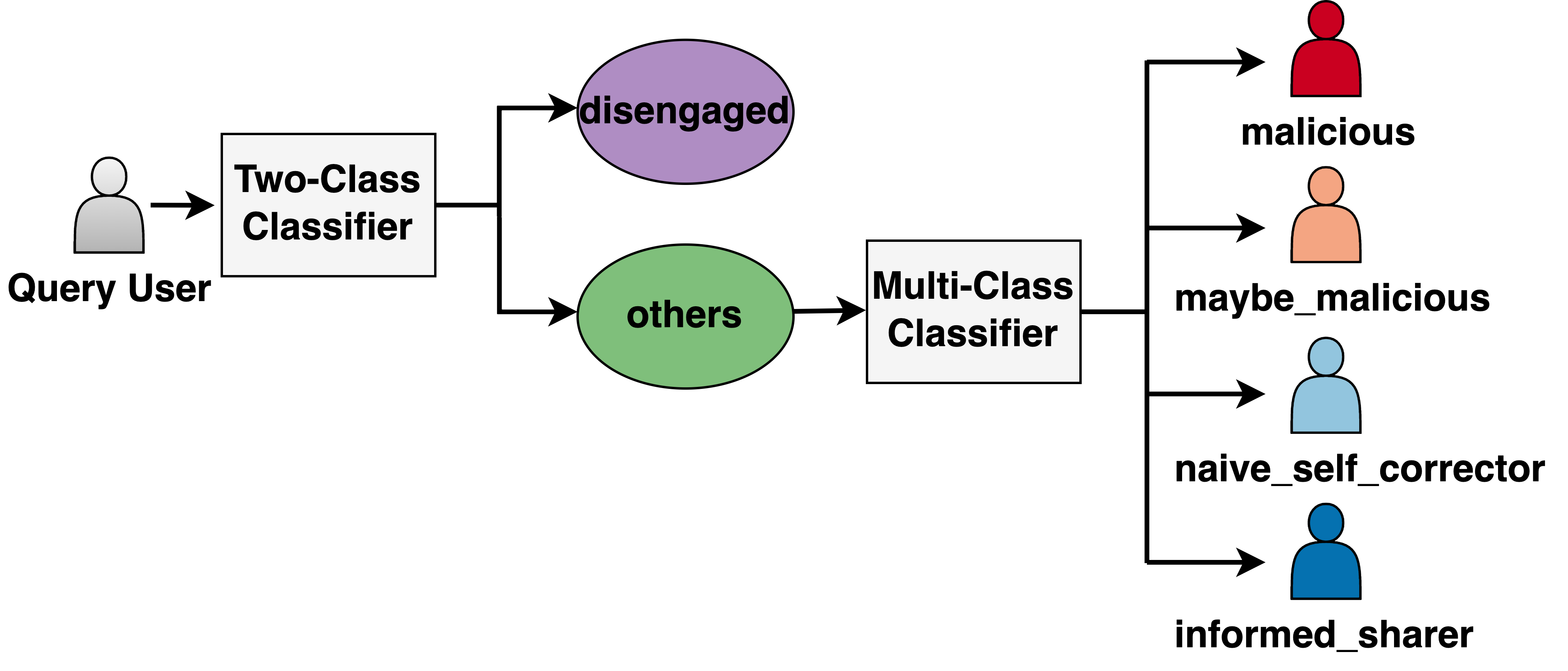}
  \caption{Two step classification overview where we use a two-class classifier to predict the \textit{disengaged} people and \textit{others} and then use a multi-class classifier to further categorize the \textit{others}.}
%   \Description{Describe the diagram.}
  \label{fig:class_overview}
\end{figure}

%% file: sections/dataset.tex
\section{Dataset}
\label{sec-dataset}
We have used the publicly available ``False and refutation information network and historical behavioral data"~\cite{rath2021false} dataset for our experiment and model evaluation. It contains misinformation and refutation related data for 10 news events (all on political topics), occuring on Twitter during 2019, identified through \texttt{altnews.in}~\cite{an:altnews}, a popular fact-checking website. For each news event, the dataset includes the single original tweet (source tweet) information for a piece of misinformation and the list of people who retweeted that misinformation along with the timestamp of the retweets. It also contains the same information for its refutation tweet. As the time of retweet is missing for news events 1 and 9, we have used data for news events 2 through 8 and 10 (total 8 news events).
\input{fig_tab_tex/tsne2}

The dataset also includes the follower-followee network information for the retweeters of the misinformation and its refutation. Since people belonging to the \textit{disengaged} category have retweeted neither of the true and false information, we had to collect their follower-followee network using Twitter API.

The following Twitter profile features of users in the follower-followee network is also included in the dataset:
\begin{itemize}
    \item Follower Count
    \item Friend (Followee) Count
    \item Statuses Count (number of tweets or retweets issued by the user)
    \item Listed Count (number of public lists the user is a member of)
    \item Verified User (True/False)
    \item Protected Account (True/False)
    \item Account Creation Time
\end{itemize}

%% file: fig_tab_tex/tsne2.tex
\begin{figure*}[t]
  \centering
  \includegraphics[width=\linewidth]{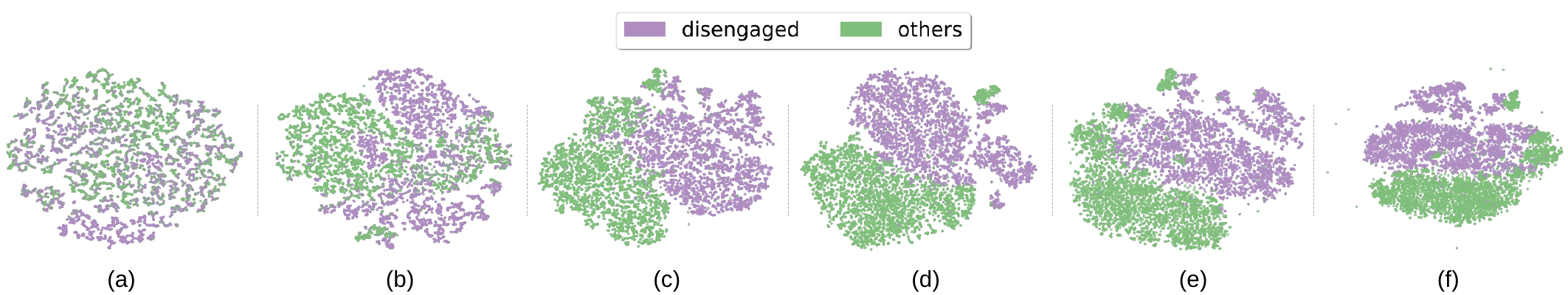}
  \caption{t-SNE plots of the learned LINE embeddings depicting two classes (\textit{disengaged} and \textit{others}) with varying feature dimensions: (a) 4d, (b) 8d, (c) 16d, (d) 32d, (e) 64d, and (f) 128d.}
%   \Description{Describe the diagram.}
% \vspace{-2mm}
  \label{fig:tsne2}
\end{figure*}

%% file: sections/experiment.tex
\section{Experimental Evaluation}
\input{fig_tab_tex/tsne4}
\input{fig_tab_tex/acc_2class}
\label{sec-expeval}

\subsection{Experiments}
\label{sec-exp}

\subsubsection{Label Creation}
\label{sec-label-creation}
First, we have labeled the people who are exposed to at least one of the eight news events' both misinformation ($r_m$) and refutation tweets ($r_f$) using the state diagram in Fig.~\ref{fig:state_diagram}. Now, a person is exposed to $r_m$ or $r_f$ when a person they follow has shared it and they see it. \textit{Twitter's data collection API does not let us collect the list of people who actually saw the tweet and the time when they saw it. This is why we assume that when a person shares $r_m$ (or $r_f$), all of their followers are exposed to it.} Similarly, we have used the following policy to set the exposure time\footnote{When Twitter or other social network platforms will use our tool, they can use the actual list of users who saw that $r_m$ (or $r_f$) as the set of exposed people along with the real exposure time.}:

\begin{itemize}
    \item We assume that when a person tweets or retweets a piece of information ($r_m$ or $r_f$), all of their followers get exposed to it. Hence, we set their exposure time equal to the (re)tweet time.
    \item If more than one followee of a person retweets the same message, then we set the exposure time of that person (to that message) to the earliest retweet time.
    \item If the user retweets the original tweet before any of the above events happens, we set the exposure time equal to the time of the user's first retweet of that message.
\end{itemize}

Comparing the sequences of exposure time and retweet time, we have been able to label people into one of the five defined categories. After labeling, we have got 1,365,929 labeled users where 99.75\% (1,362,510) of them fall into the \textit{disengaged} category and 0.25\% (3,419) of them are categorized as the other 4 classes. The number of users in these classes are: \textit{malicious}: 926, \textit{maybe\_malicious}: 222, \textit{naive\_self\_corrector}: 1,452, \textit{informed\_sharer}: 819. We can see that most of the people (around 42\%) are categorized as \textit{naive\_self\_corrector}, which indicates that most of the people who transmit misinformation, do that mistakenly. Again, the number of people in the \textit{malicious} and \textit{informed\_sharer} categories implies that the number of people in the extreme good class is almost equal, if not more, to the number of people in the extreme bad class.

\subsubsection{Feature Creation}
Now, we extract the follower-followee information of these labeled people from the dataset and construct a network. Since 99.75\% of the labeled people belong to the  \textit{disengaged} category, we randomly undersample the people belonging to this category and keep 4,059 of them for the analysis. After constructing the network, we have got 7.5M (7,548,934) nodes and 25M (25,037,335) edges. Then, we use graph embedding models LINE and PBG (as mentioned in Section~\ref{sec-emb}) to extract their network features. We have generated embeddings of different dimensions (4d, 8d, 16d, 32d, 64d, and 128d).

Next, we have normalized the embedding features. We have used the embedding features directly on various two-class classifiers for the two-class classification step (\textit{disengaged} and \textit{others}). Since the embedding features have been able to get over 99\% accuracy as discussed in Section~\ref{sec-results}, we have not included the profile features at this step. However, for the multi-class classification step, we have concatenated the normalized profile features with the learned embeddings. The profile features listed in Section~\ref{sec-dataset} were used in this step since authors in \cite{shu2018understanding} have shown that these features exhibit a significant difference between true and false information spreader groups. The boolean (True/False) features (verified user, protected account) have been converted to integers (1/0) with the account creation time being converted to normalized account age (in days).

\subsubsection{Classification}
\label{sec:exp-classification}
Since 99.75\% of the labeled people belong to the \textit{disengaged} category, and only 0.25\% belong to the other four classes, we first focus on distinguishing the \textit{disengaged} from the \textit{others} to address the rare-class problem. Then, we further distinguish among those four classes. This is why the two-step classification is used (Fig.~\ref{fig:class_overview}). We have tried different classifiers from different areas. For both the classification steps, we have used Logistic Regression as a linear classifier, \textit{k}-Nearest Neighbors algorithm (\textit{k}-NN), Support Vector  Machine (SVM), Naive Bayes as non linear classifiers, Decision Tree as tree based classifier, Random Forest (with 100 trees), Bagged Decision Tree (with base estimator SVM) as ensemble classifiers. For \textit{k}-NN, \textit{k}$=5$ has seemed to produce better results. One-vs-rest scheme has been used for Logistic Regression in the multi-class classification step. For the two-class classification step, the class distribution is almost balanced (4,059 \textit{disengaged} and 3,419 \textit{others}) after the undersampling of the \textit{disengaged} users. However, for the multi-class classification step, the class distribution is found to be imbalanced as mentioned in Section~\ref{sec-label-creation}. To account for this problem, we have set the \textit{class\_weights} parameter of the classifiers to `balanced' when available which automatically adjusts weights inversely proportional to class frequencies in the input data. For classifiers that do not have this parameter, we have used Synthetic Minority Over-sampling Technique (SMOTE)~\cite{chawla2002smote} to balance the class distribution.

\subsubsection{Baselines Used for Comparison}
As the classifier built in the two-class classification step distinguishes between the disengaged people from the others, and \textit{there is no past work which does the same,} a baseline model predicting random class (accuracy=50\%) is considered for the evaluation of this step. Regarding the evaluation of the multi-class classification step for the detection of different kinds of spreaders, no existing work does such multi-class classification; all of them do binary classification to identify true and false information spreaders. Also, their definitions of false and true information spreaders to construct the labeled data are quite different from ours. Whenever a person spreads misinformation, they label them as misinformation spreader, despite the possibility of the fact that they might not have seen the refutation and got deceived by the misinformation. We have labeled our data using the labeling mechanism described in Section~\ref{sec-foren} where people are labeled only when they are exposed to a pair of misinformation and its refutation. \textit{The results of this step are not comparable to the existing methods of detection of misinformation spreaders because of the difference in the number of classes and their definitions.} Therefore, for the multi-class classification step, two baseline models are considered: (1) Baseline 1, which predicts all samples as the majority class (\textit{naive\_self\_corrector}), and (2) Baseline 2, which predicts random class. K-fold cross-validation with K=5 and K=10 for the two-class and multi-class classification respectively has been used for evaluation purposes.

\input{fig_tab_tex/auc}
\input{fig_tab_tex/acc_weightedf1}

\subsection{Results}
\label{sec-results}

\input{fig_tab_tex/precision_recall_f1}
\input{fig_tab_tex/baseline_prf1}

% fig - shows dontcare vs others class t-SNE- discuss
Both LINE and PBG embeddings show similar results in prediction. We are using LINE embedding method to report the results since it performed faster than PBG during our experiment. Fig.~\ref{fig:tsne2} shows different t-distributed stochastic neighbor embedding (t-SNE)~\cite{van2008visualizing} plots for different dimensional LINE embeddings where separation between the \textit{disengaged} class and \textit{others} class is very clear as we increase the number of dimensions. For 4 and 8 dimensional LINE embeddings, (Fig.~\ref{fig:tsne2}a and Fig.~\ref{fig:tsne2}b), samples from both the groups are very scattered, but clusters start to appear from 16 dimension (Fig.~\ref{fig:tsne2}c). Similarly, Fig.~\ref{fig:tsne4} shows the t-SNE plots for the LINE embeddings of different dimensions for the four non \textit{disengaged} classes. As we increase the number of dimensions of the embeddings, clusters start forming, i.e., people from the \textit{malicious} class have formed a cluster on the right and top part of Fig.~\ref{fig:tsne4}f whereas people from the \textit{informed\_sharer} class have formed a cluster on the bottom part of that Figure.

\subsubsection{Two-Class Classifier}

Table~\ref{tab:acc_2class} shows the performance of the two-class classifiers using LINE embeddings with 128 dimensions. After using embeddings from different dimensions, we have observed that Logistic Regression and Bagged Decision Tree have consistently performed better (accuracy greater than 98\%) than other classifiers when the number of dimensions is above 16. Logistic Regression achieves 99.251\% accuracy for 128-dimensional LINE embeddings which outperforms the baseline model of accuracy = 50\%. Fig.~\ref{fig:auc} depicts the Receiver Operating Characteristic (ROC) curve for 5-fold cross-validation which demonstrates the efficacy of the model.

\subsubsection{Multi-Class Classifier}
Table~\ref{tab:acc_weightedf1} reports the accuracy and the weighted F1 score of the multi-class classification step using 128 dimensional LINE embeddings whereas Fig.~\ref{fig:precision_recall_f1} depicts their precision, recall and F1 score using a bar plot. \textit{k}-NN and bagged decision tree have consistently shown to produce better output than other classifiers where bagged decision tree has performed slightly better than \textit{k}-NN. For example, the precision for the \textit{malicious} class using \textit{k}-NN is 75.812\% and using bagged decision tree is 77.446\%. The recall for this class is about the same ($\sim$75\%) for both classifiers. The precision for the \textit{maybe\_malicious}, \textit{naive\_self\_corrector} and \textit{informed\_sharer} classes are 92.078\%, 64.246\%, 69.073\% using bagged decision tree, respectively and 54.57\%, 59.893\%, 60.327\% using \textit{k}-NN, respectively. Moreover, Bagged Decision Tree has achieved an accuracy score of 73.637\% and a weighted F1 score of 72.215\%. All of these numbers demonstrate the better prediction capability of these models than the baseline models shown in Table~\ref{tab:baseline_prf1}.
After comparing the performance of the models using embeddings of different dimensions, we have observed improvement in the performance as we have increased the number of dimensions\footnote{The code used in this analysis is available at \url{https://github.com/eunakhan/behavioral-forensics}.}.

%% file: fig_tab_tex/tsne4.tex
\begin{figure*}[t]
  \centering
  \includegraphics[width=\linewidth]{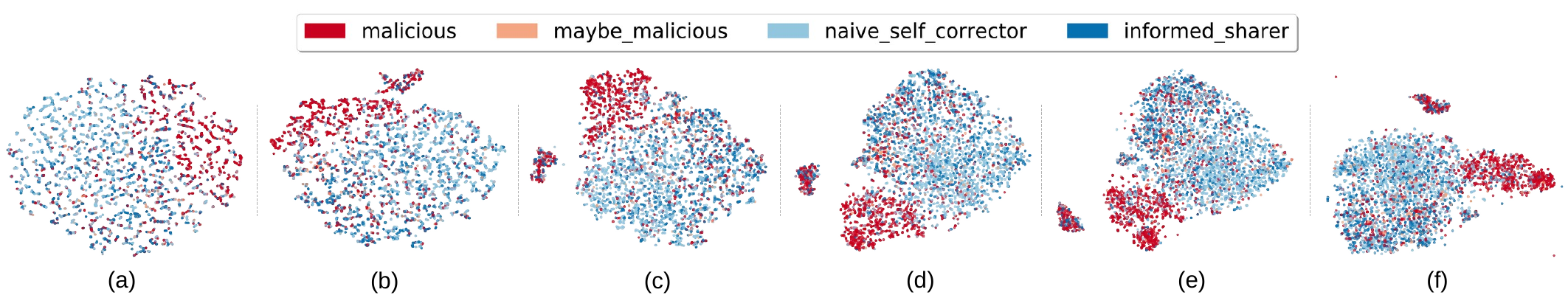}
  \caption{t-SNE plots of the learned LINE embeddings depicting the four non \textit{disengaged} classes  with varying feature dimensions: (a) 4d, (b) 8d, (c) 16d, (d) 32d, (e) 64d, and (f) 128d. (Best viewed at 300\% zoom)}
  % \vspace{-4mm}
%   \Description{Describe the diagram.}
  \label{fig:tsne4}
\end{figure*}

%% file: fig_tab_tex/acc_2class.tex
\begin{table*}[t]
  \caption{Accuracy (\%) of different machine learning models for the two-class classification step using 5-fold cross-validation.}
  % \vspace{2mm}
  \label{tab:acc_2class}
  % \begin{adjustbox}{max width=\textwidth}
  \begin{tabular}{lccccccc}
    \toprule
    \textbf{Classifier} &   \textit{k}-NN   &  Logistic Regression &  Naive Bayes & Decision Tree & Random Forest & SVM & Bagged Decision Tree \\
    \midrule
    \textbf{Accuracy}     & 95.318 & \textbf{99.251} & 81.236 & 86.936 & 97.794 & 98.957 & \textbf{99.157} \\
  \bottomrule
\end{tabular}
% \end{adjustbox}
% \vspace{-4mm}
\end{table*}

%% file: fig_tab_tex/auc.tex
\begin{figure}[t]
  \centering
  \includegraphics[width=\linewidth]{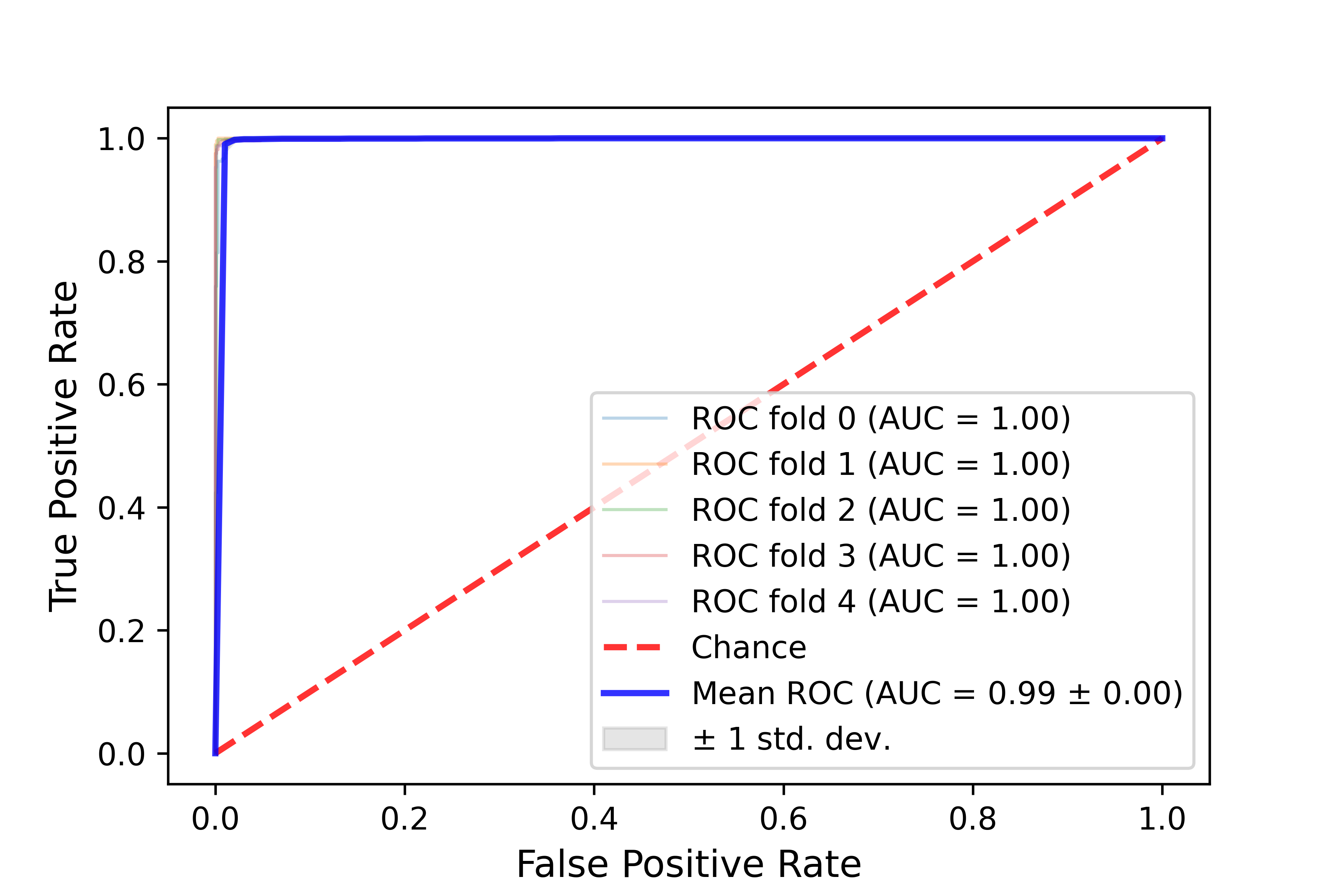}
  \caption{ROC curve for the two-class classification step using 5-fold cross-validation.}
%   \Description{Describe the diagram.}
% \vspace{-4mm}
  \label{fig:auc}
\end{figure}

%% file: fig_tab_tex/acc_weightedf1.tex
\begin{table}
  \caption{Accuracy (\%) and weighted F1 score (\%) of different machine learning models along with baseline models for the multi-class classification step using 10-fold cross-validation.}
  % \vspace{2mm}
  \label{tab:acc_weightedf1}
  % \begin{adjustbox}{max width=\textwidth}
  \begin{tabular}{lcc}
    \toprule
    \textbf{Classifier} &    \textbf{Accuracy}  &   \textbf{Weighted F1}\\
    \midrule
    Baseline 1 & 41.329 & 24.172 \\
    Baseline 2 & 25.108 & 26.988 \\
    \textit{k}-NN & 61.412 & 53.095\\
    Logistic Regression & 48.273 & 47.451\\
    Naive Bayes & 53.328 & 51.926\\
    Decision Tree & 41.758 & 41.247\\
    Random Forest & 53.546 & 45.285\\
    SVM & 52.239 & 50.783\\
    Bagged Decision Tree & \textbf{73.637} & \textbf{72.215}\\
  \bottomrule
\end{tabular}
% \end{adjustbox}
% \vspace{-4mm}
\end{table}

%% file: fig_tab_tex/precision_recall_f1.tex
\begin{figure*}[t]
  \centering
  \includegraphics[width=\linewidth]{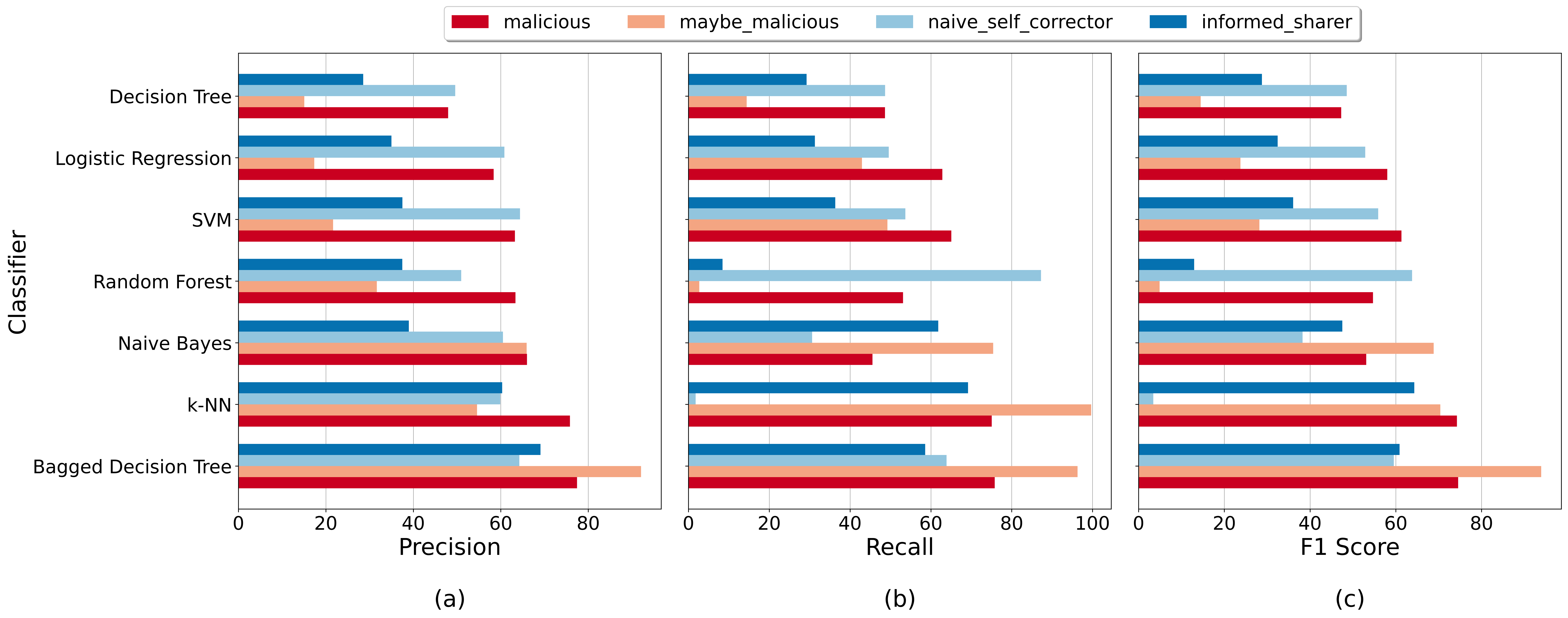}
  \caption{Horizontal bar plots showing the (a) precision, (b) recall, and (c) F1 score of four different non \textit{disengaged} classes using different machine learning models for the multi-class classification step. All the values are in percentages.}
%   \Description{Describe the diagram.}
% \vspace{-4mm}
  \label{fig:precision_recall_f1}
\end{figure*}

%% file: fig_tab_tex/baseline_prf1.tex
\begin{table*}
  \caption{Precision, recall and F1 score of four different non \textit{disengaged} classes using the baseline models for the multi-class classification step. All the values are in percentages. Note that while predicting all samples as the majority class, baseline 1 produces undefined precision and F1 score for the non-majority classes as expected, which are represented by `-' in the table.}
  \vspace{2mm}
  \label{tab:baseline_prf1}
  \centering
  \begin{tabular}{lcccccc}
    \toprule
    \multirow{2}{*}{\textbf{Class Categories}} &  \multicolumn{3}{c}{\textbf{Baseline 1}}  & \multicolumn{3}{c}{\textbf{Baseline 2}}\\
    \cmidrule(lr){2-4}\cmidrule(lr){5-7}
     & \multicolumn{1}{c}{\textbf{Precision}} & \multicolumn{1}{c}{\textbf{Recall}} & \multicolumn{1}{c}{\textbf{F1 Score}} & \multicolumn{1}{c}{\textbf{Precision}} & \multicolumn{1}{c}{\textbf{Recall}} & \multicolumn{1}{c}{\textbf{F1 Score}}\\
    \midrule
    malicious & - & 0 & - & 25.795 & 23.397 & 24.537\\
    maybe\_malicious & - & 0 & - & 6.388 & 23.423 & 10.038\\
    naive\_self\_corrector & 41.329 & 100 & 58.486 & 42.430 & 26.229 & 32.418\\
    informed\_sharer & - & 0 & - & 23.954 & 25.511 & 24.708\\
  \bottomrule
\end{tabular}
% \vspace{-4mm}
\end{table*}

%% file: sections/discussion.tex
\section{Discussion}
\label{sec-discussion}
This paper proposes a \textit{behavioral forensics} model to identify different types of misinformation, disinformation and refutation spreaders. Our experimental results show the efficacy of our proposed model. We have used different dimensions of embeddings and observed that increasing the number of dimensions improves the performance of the model initially but this improvement slows down as we reach 64d. It is interesting to note from the t-SNE plots (Fig.~\ref{fig:tsne2} and Fig.~\ref{fig:tsne4}) that people from different groups can form different clusters when only network embedding features are used. It signifies the importance of network structure in spreader detection and the classification results prove that network embedding features can be used for people's behavioral aspect, such as spreading behavior prediction. However, the metric which should be used for model selection and tuning depends on the mitigation techniques used to fight misinformation dissemination. For instance, if \textit{malicious} people are decided to be banned, then we will want to put an emphasis on its precision since we do not want to ban any good account. On the other hand, if treating the followers of the \textit{malicious} people with refutation is taken as a preventive measure, then we might want to focus on its recall, and select and tune the model based on that. If both measures are taken, then the F1 score has to be maximized. Again, if people from the \textit{naive\_self\_corrector} and \textit{informed\_sharer} classes are provided with refutation to facilitate the spread of true information, then their recall should be maximized. These peer corrections can reduce misinformation sharing~\cite{margolin2018political} and mitigate misperceptions among the community seeing the interaction~\cite{vraga2020correction}. 

%% file: sections/conclusion.tex
\section{Conclusion \& Future Work}
\label{sec-conclusion}
This research proposes a novel way to define and classify different kinds of spreaders by introducing a multi-class behavior-based labeling mechanism. Exposure to both true and false information is required for labeling a person and their reactions to those messages are observed to label them. It is also shown that people from different categories can form different clusters when network embedding features are used and those features can be used to train machine learning models to predict their labels. In this way, our proposed model is able to predict the category of a person without having behavioral history of actions. The results show the efficacy of our proposed model. Our model can be applied to any social network to fight misinformation spread. Possible future work includes updating people's labels when they exhibit a new or different behavior from the past. It will enable the model to be adaptive to people's behavior change and be more dynamic. Another possible extension to this work can be observing how the categories of people change when they are exposed to misinformation from different topics.